\documentclass{aa}
\usepackage{graphicx,natbib}
\bibpunct{(}{)}{;}{a}{}{,}
\usepackage{txfonts}
\begin{document}
\title{Plasma diagnostic in eruptive prominences from SDO/AIA observations at 304~\AA}
\author{Nicolas Labrosse \and Kristopher McGlinchey}
\institute{SUPA, School of Physics and Astronomy, University of Glasgow, Glasgow, G12 8QQ, Scotland\\
\email{Nicolas.Labrosse@glasgow.ac.uk}
}
\abstract
{Theoretical calculations have shown that when solar prominences move away from the surface of the Sun, their radiative output is affected via the Doppler dimming or brightening effects.
}
{In this paper we ask whether observational signatures of the changes in the radiative output of eruptive prominences can be found in EUV (extreme ultraviolet) observations of the first resonance line of ionised helium at 304~\AA. We also investigate whether these observations can be used to perform a diagnostic of the plasma of the eruptive prominence.
}
{We first look for suitable events in the SDO/AIA database. The variation of intensity of arbitrarily selected features in the 304 channel is studied as a function of velocity in the plane of the sky. These results are then compared with new non-LTE radiative transfer calculations of the intensity of the \ion{He}{ii} 304 resonance line.}
{We find that observations of intensities in various parts of the four eruptive prominences studied here are sometimes consistent with the Doppler dimming effect on the \ion{He}{ii} 304~\AA\ line. However, in some cases, one observes an increase in intensity in the 304 channel with velocity, in contradiction to what is expected from the Doppler dimming effect alone. The use of the non-LTE models allows us to explain the different behaviour of the intensity by changes in the plasma parameters inside the prominence, in particular the column mass of the plasma and its temperature.}
{The non-LTE models used here are more realistic than what was used in previous calculations. They are able to reproduce qualitatively the range of observations from SDO/AIA analysed in this study. Thanks to non-LTE modelling, we can infer the plasma parameters in eruptive prominences from SDO/AIA observations at 304~\AA.}
\keywords{Sun: filaments, prominences -- Line: formation -- Radiative transfer -- Sun: corona}
\authorrunning{Labrosse \& McGlinchey}
\titlerunning{Diagnostics of eruptive prominences with SDO/AIA at 304~\AA}
\maketitle

\section{Introduction}
Solar prominences are strongly sensitive to the radiation coming from the solar disc, which creates
conditions far from Local Thermodynamic Equilibrium \citep[LTE -- see][]{1963PASJ...15..122H}. They are also optically thick in
several spectral lines. To model the emitted radiation it is necessary to use non-LTE radiative
transfer codes.

{Non-LTE radiative transfer} calculations have shown that when the prominences move away from the
surface of the Sun, for instance when they become active or eruptive, their radiative output is
affected \citep[see review by][and references therein]{2010SSRv..151..243L}. If this can be observed then this opens a new door to
diagnose the prominence plasma in the eruption phase. This predicted change in the radiation output
is due to the so-called Doppler dimming / brightening effect {\citep{1970SoPh...14..147H}}, and is widely used to diagnose the
solar wind speed from the observation of resonance lines {\citep[see e.g.,][]{1982ApJ...256..263K}}.

In this study we investigate whether observational signatures of the changes in the radiative output
from eruptive prominences can be found in EUV observations, in particular in the 304~\AA\ line
emitted by ionised helium. This line has been shown to be very sensitive to the radial motion of the
prominence, because it is mostly formed by scattering of the incident radiation coming from the
Sun \citep{2007A&A...463.1171L,2008AnGeo..26.2961L}. {The present work is therefore} an attempt to develop an observational answer to
the questions raised by the aforementioned theoretical studies.
To do this, we look for suitable events in the SDO/AIA database, and measure the intensities emitted by these prominences. AIA \citep[Atmospheric Imaging Assembly,][]{2011SoPh..tmp..115L} is a multi-band imager including the He II 304~\AA\ channel and is one of three instruments embarked on the Solar Dynamics Observatory {(SDO)}. These observations are then compared with a new grid of {1D} non-LTE models.
The main objective of this research is to characterise the variation in the radiation of active and eruptive
prominences as they move away from the solar surface and therefore receive a variable
incident radiation. The small sample of well-observed prominence eruptions
selected for this study allows us to answer the questions posed by the theoretical studies, namely 1) can we
observe the predicted intensity variations, and 2) if yes can we use this effect to diagnose the
prominence plasma? These objectives are primarily related to a number of open issues in
solar prominence physics \citep{2010SSRv..151..243L}.

The paper is organised as follows: observations are described in Sect.~\ref{s:obs}. The non-LTE models are described in Sect.~\ref{s:models}. In Sect.~\ref{s:discuss} we discuss our results before the conclusions in Sect.~\ref{s:concl}.

\section{Observations}\label{s:obs}

Launched in February 2010 by NASA,
{AIA}
{onboard}
{SDO} is the most advanced
solar imager to date. It consists of 4 dual-channel telescopes with filters {centered} at 94~\AA, 131~\AA, 171~\AA, 193~\AA, 211~\AA, 304~\AA, 335~\AA\ and 1600~\AA. Here we use observations at 304~\AA, corresponding primarily to \ion{He}{ii} emission coming from the chromosphere and the transition region at a temperature of about $\log T=4.7$. In that passband, however, lies another line at 303.3~\AA\ emitted by \ion{Si}{xi} (see Sect.~\ref{s:contrib}).
The main advantage of AIA over previous solar imagers is both its high spatial and temporal resolutions. The apparatus uses a 4096$\times$4096 CCD chip with a pixel size of 0.6\arcsec\ (level 1.5 data), allowing it to probe into the fine structures of the Sun. AIA takes an image approximately every 12 seconds at each wavelength, allowing it to closely track even the
most rapidly evolving features -- which is useful for our investigation.

We select four events in the SDO/AIA database. The main selection criteria are that the eruptions must be well developed and last long enough to facilitate manual feature tracking.
We choose four events exhibiting different morphologies.
In each case, we produce a movie {(available on request)} from the level 1.5 data and follow individual features. Intensities are averaged over nine pixels to produce the intensity plots shown below {(Figs~2--5)}. All intensities are normalised by the exposure time (2.9~s). {In order} to make the comparison with the theoretical {intensities} presented in Sect.~\ref{s:models}, the {observed} intensities are normalised by the intensity corresponding to the lowest velocity. {Hence all intensity values greater than 1 indicate that a brightening of the prominence with velocity, while intensity values lower than 1 indicate a dimming of the prominence with velocity.}
{The variation of the relative intensity as a function of velocity shown in the bottom panels of Figs~2--5 has been plotted with the same axis scales in intensity and velocity for the four events in order to ease the comparison between each of them. In Fig.~\ref{f:130610} an inset has been added to show this intensity variation on a more appropriate scale for this specific event.}

{Figure~\ref{f:veltime} shows the evolution of the velocity of selected features in four distinct eruptive prominences as a function of time elapsed since the beginning of tracking.
\begin{figure}
\resizebox{\hsize}{!}{\includegraphics[angle=90]{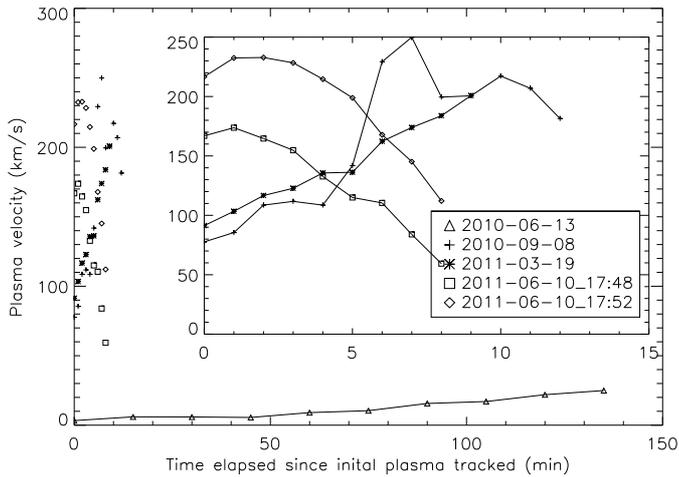}}
\caption{{Temporal variation of the velocity of the selected features for the four prominence eruptions studied. {The inset shows this velocity variation in more details for fast-evolving events.}}}
\label{f:veltime}
\end{figure}
This gives an overview of the dynamics of the selected events. While most of them evolve quite rapidly (reaching velocities larger than 100~km s$^{-1}$ within a few minutes), the prominence eruption observed on 13 June 2010 is much less dynamic, with velocities of a few tens km s$^{-1}$ after 1--2 hours.}

The {following} subsections present {the details  of individual events used in this  work, and} the intensity variations as a function of velocity for the selected features.

\subsection{Prominence eruption on 2010-06-13 at 00:00}
This moderate-sized prominence eruption takes place on the 13th June 2010 at 00:00 on the North-West limb. For the first few hours of the event there is a slow rise phase. From around 03:00
onwards however the top part of the prominence begins to detach itself from the base as the motion becomes more prominent. At
around 05:00 there is a clear detachment of the structure and this then moves outward. This rise continues for several hours until it suddenly stretches before fading away. The base of the prominence rises {by} a small amount during this {time}, and then blows away and disappears from view.
This eruption has been recently studied by \cite{2011A&A...533L...1R} who investigated the thermal and magnetic structure of the coronal cavity located above the prominence.

\begin{figure}
\resizebox{\hsize}{!}{\includegraphics{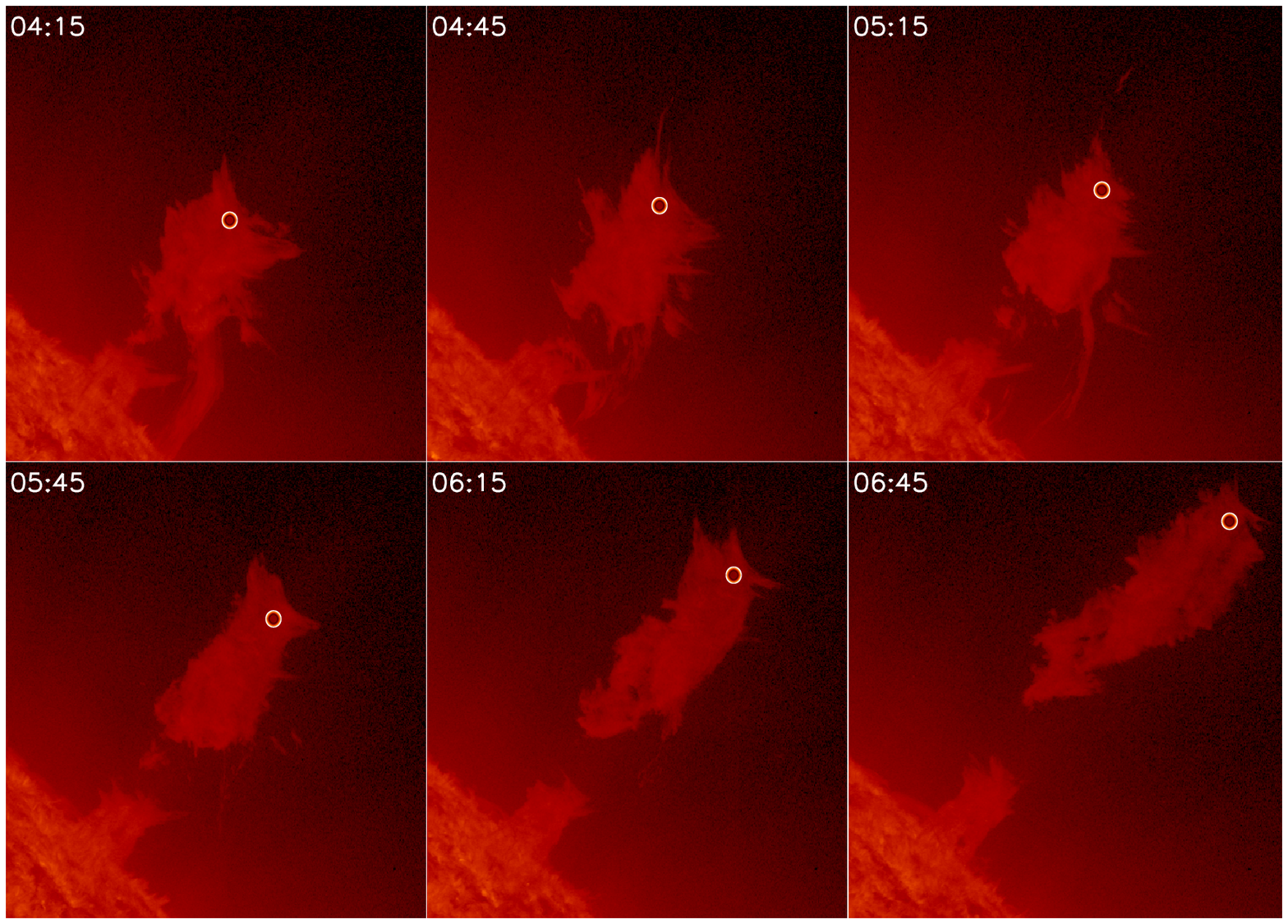}}
\resizebox{\hsize}{!}{\includegraphics[angle=90]{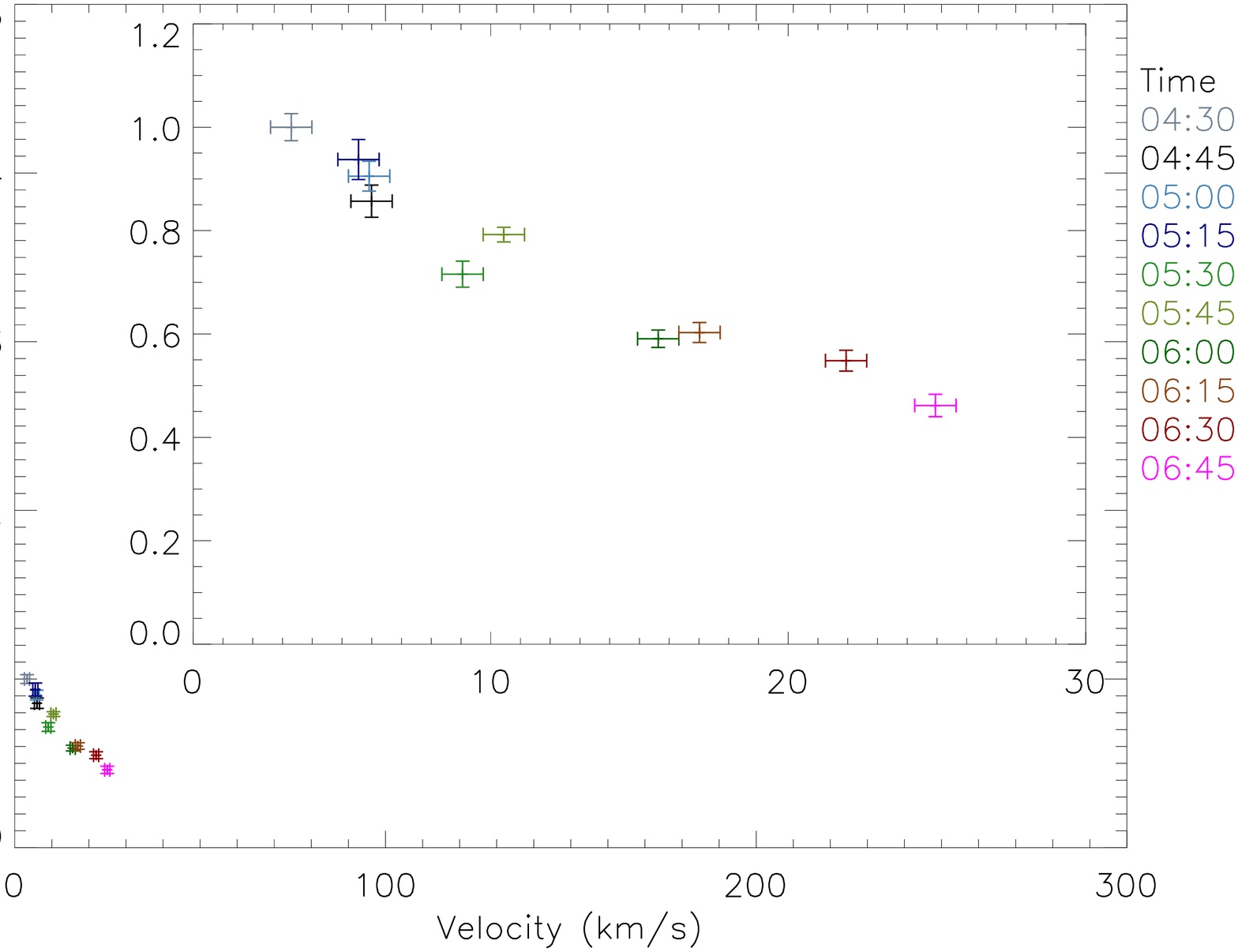}}
\caption{\textit{Top}: Evolution of the 2010-06-13 prominence eruption. The circle marks the part of the prominence which was tracked and used to calculate the intensities. The field of view in the images is 300\arcsec$\times$300\arcsec. \textit{Bottom}: Variation of relative intensity as a function of velocity in the plane of the sky. {Intensities are normalised by the intensity corresponding to the lowest velocity. {The inset shows the relative intensity variation with axis scales specific to this event. The time evolution is coded in color (with the time-scale indicated on the right of the plot) in the online version of the figure.}}}
\label{f:130610}
\end{figure}

We start our feature tracking at 04:15~UT in the middle part of the prominence body, but close to the edge to the coronal cavity. Differing from other events studied in this paper by being a substantially longer lasting, slower event, images were taken at 15 minute intervals. The tracking was taken over 11 images, ending at 06:45, just before the detached plasma is stretched.
Figure~\ref{f:130610} shows six images of the evolution of the eruptive prominence and the feature which was tracked, and the variation of relative intensity (normalised to the intensity at the lowest velocity over the tracking period) as a function of velocity. {The circles on the images highlight the pixels used in calculating the intensities.}
The velocities are much lower in this event than in the others studied
{in this paper.}

This
event shows a decrease of the observed intensity of the \ion{He}{ii} 304 channel with an increasing velocity.
The trend of an intensity decrease with increasing velocity
{is reminiscent of}
the Doppler dimming effect as illustrated by the theoretical calculations of \cite{2007A&A...463.1171L,2008AnGeo..26.2961L}.
{However, these theoretical calculations were done by varying only the radial velocity of the moving prominence, and keeping all the plasma parameters constant throughout the activation / eruption. In fact, we do not exclude that the physical parameters of the prominence plasma may vary during the eruption}.

\subsection{Prominence eruption on 2010-09-08 at 22:30}

\begin{figure}
\resizebox{\hsize}{!}{\includegraphics{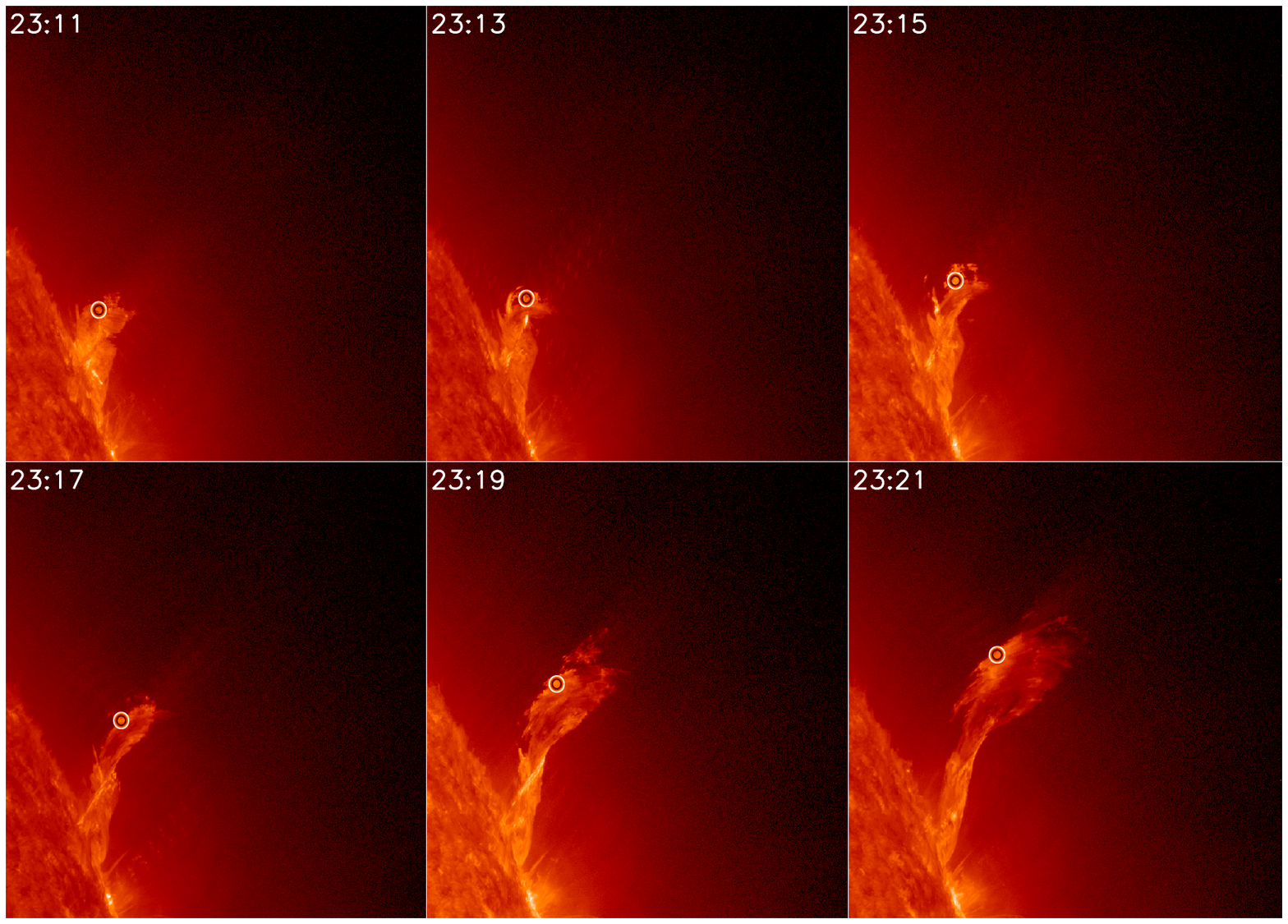}}
\resizebox{\hsize}{!}{\includegraphics[angle=90]{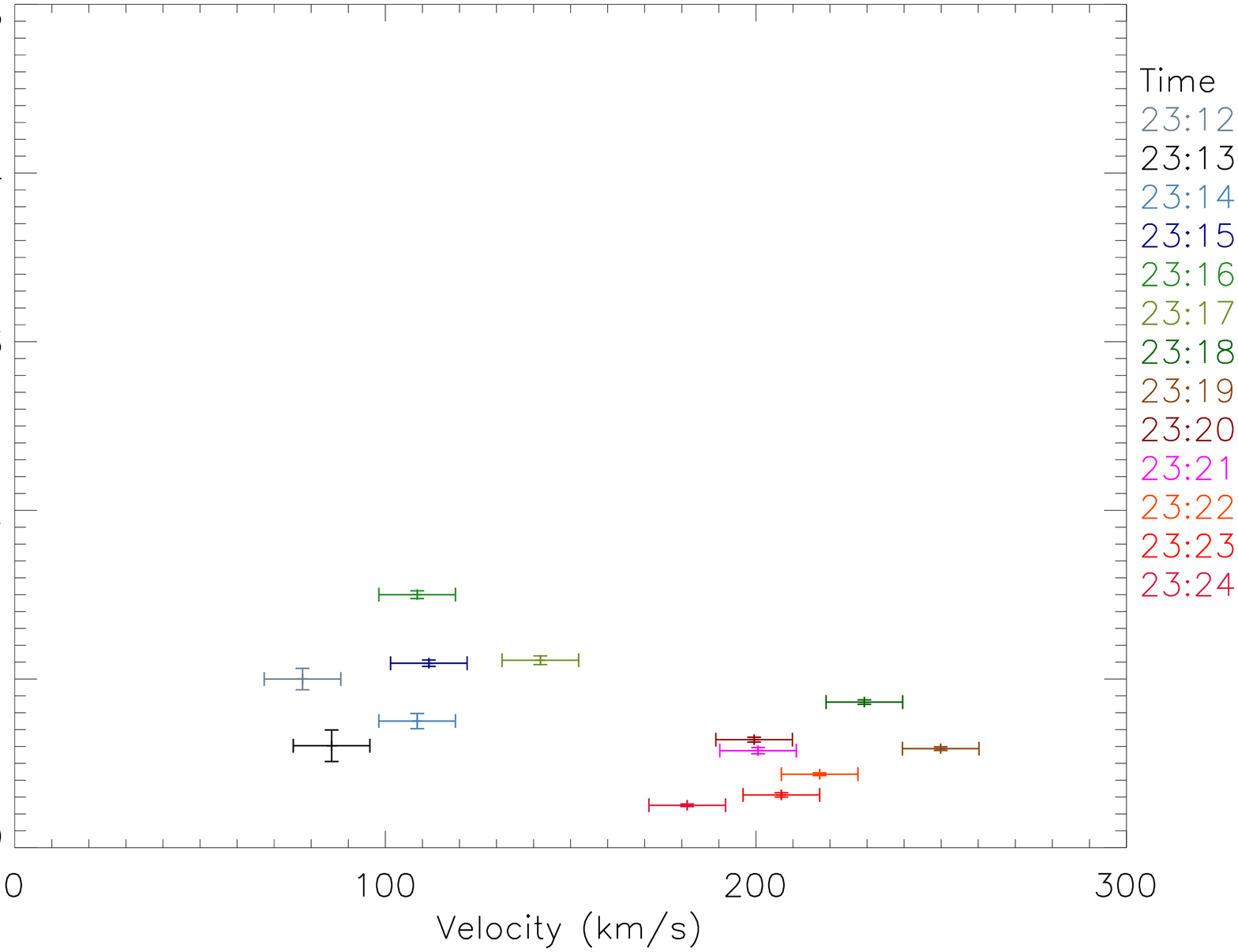}}
\caption{{Same as Fig.~\ref{f:130610}} for the 2010-09-08 prominence eruption. The field of view in the images {(top panel)} is 450\arcsec$\times$450\arcsec.}
\label{f:080910}
\end{figure}

This prominence eruption took place on September 8th, 2010. The eruption began at approximately 22:30~UT at the north-west limb. It was associated with a GOES Class B3.4 flare near AR~11105.
Any appreciable outward motion of the plasma did not occur until approximately 23:08.
The tracking starts at 23:11 over 14 frames until 23:24 at a one minute cadence.
Figure~\ref{f:080910} shows {the same information as Figure~\ref{f:130610} only for the 2010-09-08 event}.
{Here again,} the main feature of the relative intensity \textit{vs} velocity graph is the decrease in intensity with increasing velocity.

\subsection{Prominence eruption on 2011-03-19 at 11:30}

\begin{figure}
\resizebox{\hsize}{!}{\includegraphics{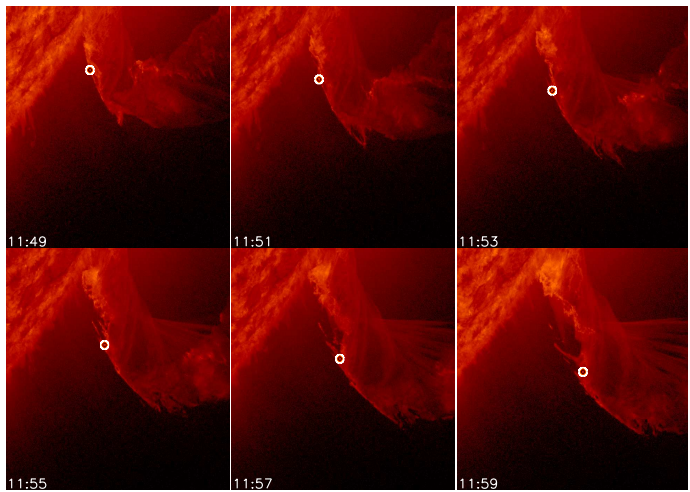}}
\resizebox{\hsize}{!}{\includegraphics[angle=90]{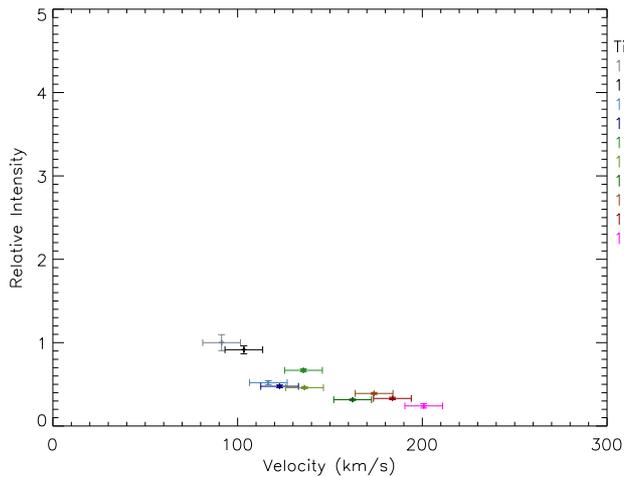}}
\caption{{Same as Fig.~\ref{f:130610}} for the 2011-03-19 prominence eruption.  The field of view in the images {(top panel)} is  390\arcsec$\times$390\arcsec. }
\label{f:190311}
\end{figure}

This other prominence eruption took place on the 19th March 2011. Beginning at 11:10~UT, AIA observed a large kinking loop eruption from AR~11171.
Located in the south west quadrant of the Sun, this eruption exhibits a strong sweeping motion as it moves further south. Helical motions are visible in the prominence legs during the eruption.
We start the manual tracking at 11:49 of a bright patch of plasma on the ridge of the prominence over 11 images ending at 11:59. Figure~\ref{f:190311} shows {the same information as Figure~\ref{f:130610} only for the 2011-03-19 event}.
It highlights what was already visible in Figs.~{\ref{f:130610} and} \ref{f:080910}, namely that a decrease of intensity with velocity is observed during the eruption.

\subsection{Prominence eruption on 2011-06-10 at 17:43}\label{s:100611}

This prominence eruption took place on the 10th June 2011 and was associated to a C2.9 flare in AR~11227. First appearing as a compact flaring region on the south west limb, the prominence soon developed into a very tight looped structure at around 17:49, very different to the other cases explored here.
This is a good example of a so-called failed eruption, with most of the plasma falling back towards the surface of the Sun.

\begin{figure}
\resizebox{\hsize}{!}{\includegraphics{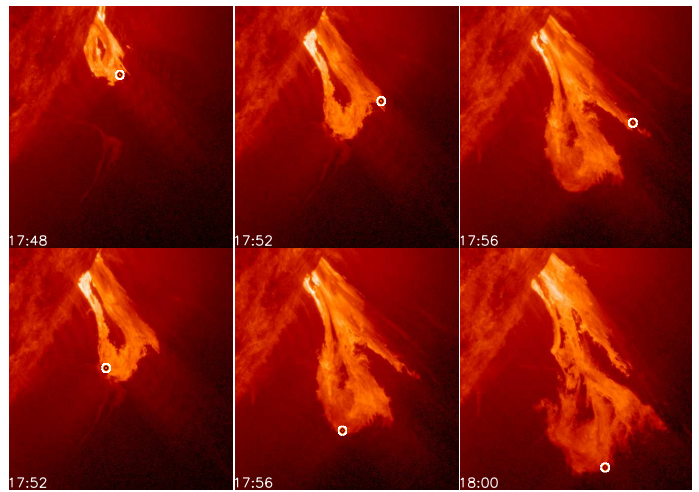}}
\resizebox{\hsize}{!}{\includegraphics[angle=90]{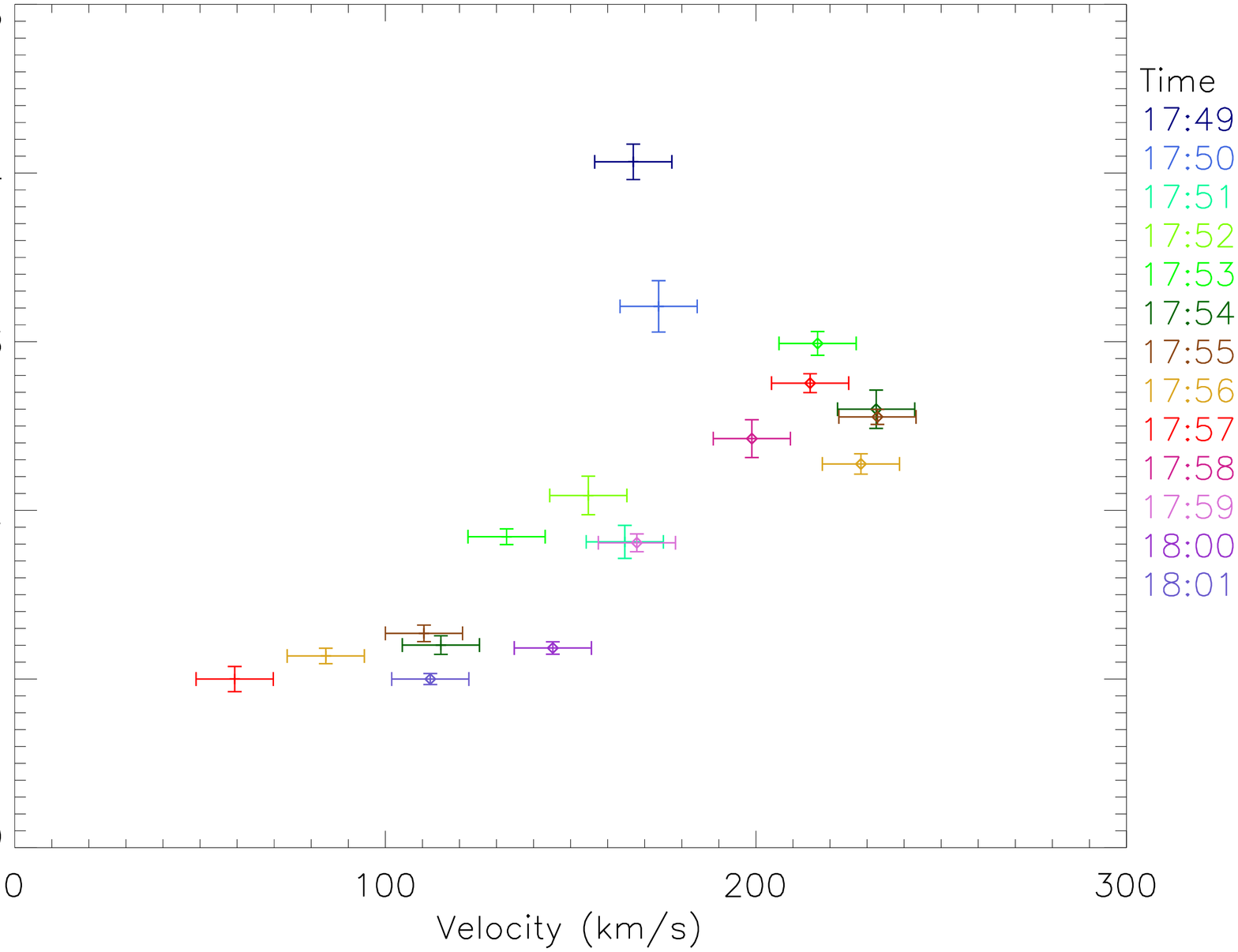}}
\caption{ {Same as Fig.~\ref{f:130610} for the 2011-06-10 prominence eruption}. The first tracked feature is indicated on the first three images in the sequence, while the last three images show the second tracked feature. The field of view in the images is 276\arcsec$\times$276\arcsec. {For the intensity-velocity graph the cross symbols correspond to the event beginning at 17:48  and the diamond symbols correspond to the 17:52 event.}}
\label{f:100611}
\end{figure}

For this particular prominence, two parts were tracked. The first feature-tracking started at 17:48 on the part of the loop which evolves into the jet of plasma. The feature was tracked over 10 images, ending at 17:57. The second blob of plasma is chosen in a different part of the prominence. Tracking started at 17:52 for 10 images, ending at 18:01.
Figure~\ref{f:100611} shows {the same information as Figure~\ref{f:130610} only for the 2011-06-10 event. In this case however the first tracked feature is indicated in the first three images in the sequence, while the last three images show the second feature}.

This case is interesting, since instead of having a decrease of intensity with velocity as was expected from the simple theoretical models of \cite{2007A&A...463.1171L,2008AnGeo..26.2961L}, and supported by the observations shown in previous sections, one now observes {an increase in intensity with velocity}.
{The two independent parts of plasma tracked show a differing intensity profile from each other}.
In Sect.~\ref{s:models} we discuss in more details what are the possible causes for this increasing intensity with velocity, but it is likely that this is an effect of a variation of the physical conditions of the prominence plasma during the eruption, which was not taken into account in the isothermal models of \cite{2007A&A...463.1171L} or in the prominence models including a transition region \citep{2008AnGeo..26.2961L}.

\section{Modelling}\label{s:models}

The motivation to construct new non-LTE models to study the Doppler dimming effect on the \ion{He}{ii} 304~\AA\ resonance line is two-fold. First, we want to test if these models are able to reproduce the observations presented in Sect.~\ref{s:obs}, at least on a qualitative basis. Second, the calculations presented in \cite{2007A&A...463.1171L,2008AnGeo..26.2961L} were carried out for simplified models which allowed these authors to understand the basic Doppler dimming mechanisms and their effects on the spectral lines produced by neutral and ionised helium. These calculations showed that the resonance lines of neutral, and to a greater extent, ionised helium, are sensitive to the Doppler dimming effect. This is because resonance scattering plays such an important role in the formation of these lines. However, the physical conditions inside their eruptive prominence models were not very realistic, since all the plasma parameters were kept constant during the eruption as the structure was moving upwards. In reality, the physical conditions within the eruptive structures may vary, depending mainly on the driving mechanisms of the eruption.

{In the following we describe our new prominence models where we relax all the model input parameters, allowing them to vary in order to study how the emergent intensities are modified when both the velocity and the other plasma parameters change during the eruption.}

\subsection{Description of the models}

\begin{figure}
\resizebox{\hsize}{!}{\includegraphics{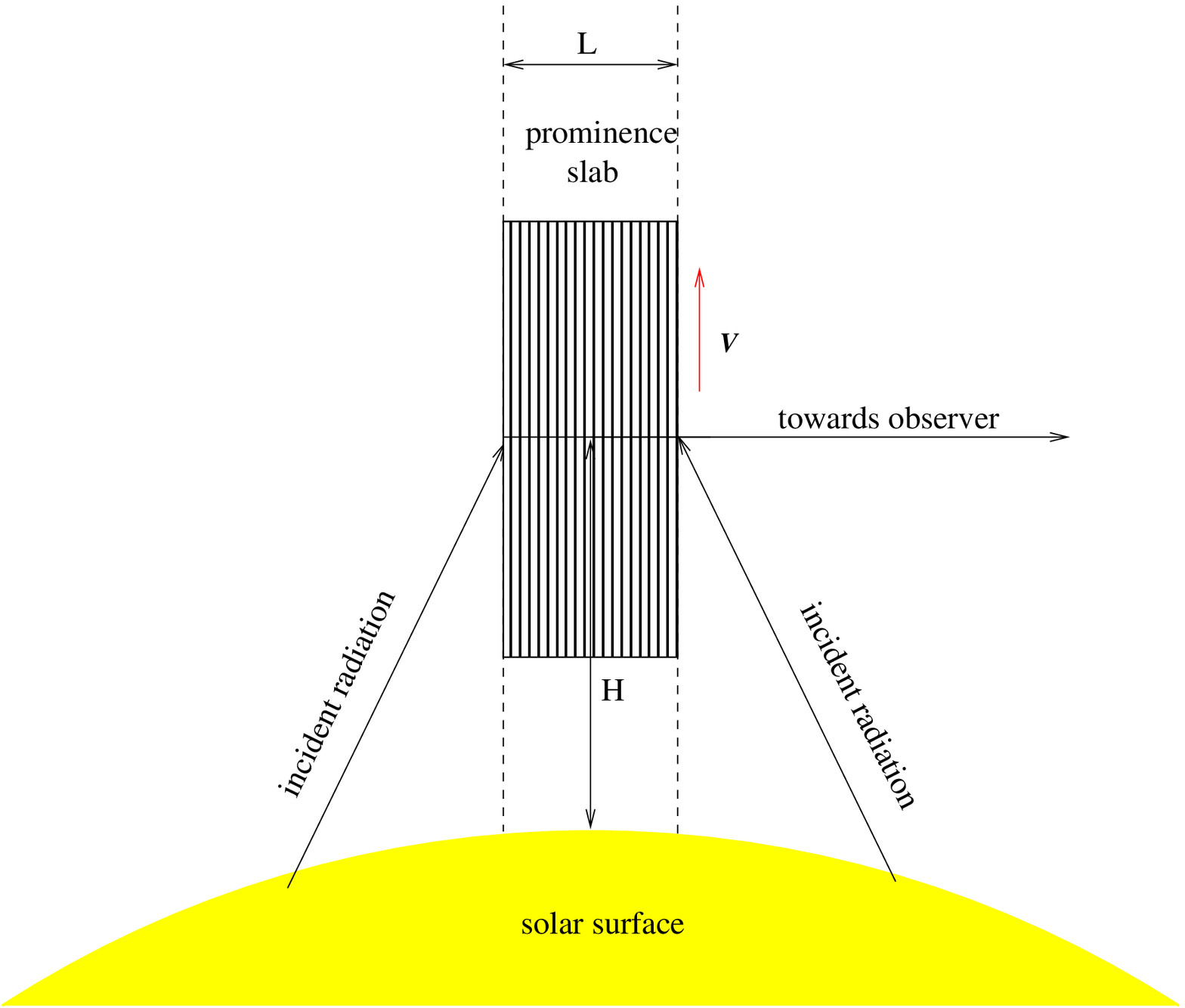}}
\caption{{Schematic representation of the 1D plane-parallel prominence slab of geometrical width $L$, altitude $H$ above the solar surface, moving radially at a velocity $V$.}}
\label{f:slab}
\end{figure}

The details of the computations are the same as in \cite{2007A&A...463.1171L,2008AnGeo..26.2961L}.
Our eruptive prominence is represented by a 1D plane-parallel slab standing vertically above the solar surface {(see Fig.~\ref{f:slab})}.
The model input parameters are the radial velocity {($V$)} at which the prominence slab is moving, the temperatures and pressures at the centre of the slab and at the boundary with the corona, the steepness $\gamma$ of the temperature gradient in the prominence-to-corona transition region (PCTR), the total column mass {(or equivalently the slab width $L$)}, the microturbulent velocity, and the altitude of the slab {$H$ (\emph{viz.,} the height of the line-of-sight above the limb where it intersects the plane of the sky)}.
The temperature and pressure profiles are taken from \cite{1999A&A...349..974A}.
\begin{figure*}
\resizebox{\hsize}{!}{\includegraphics[angle=90]{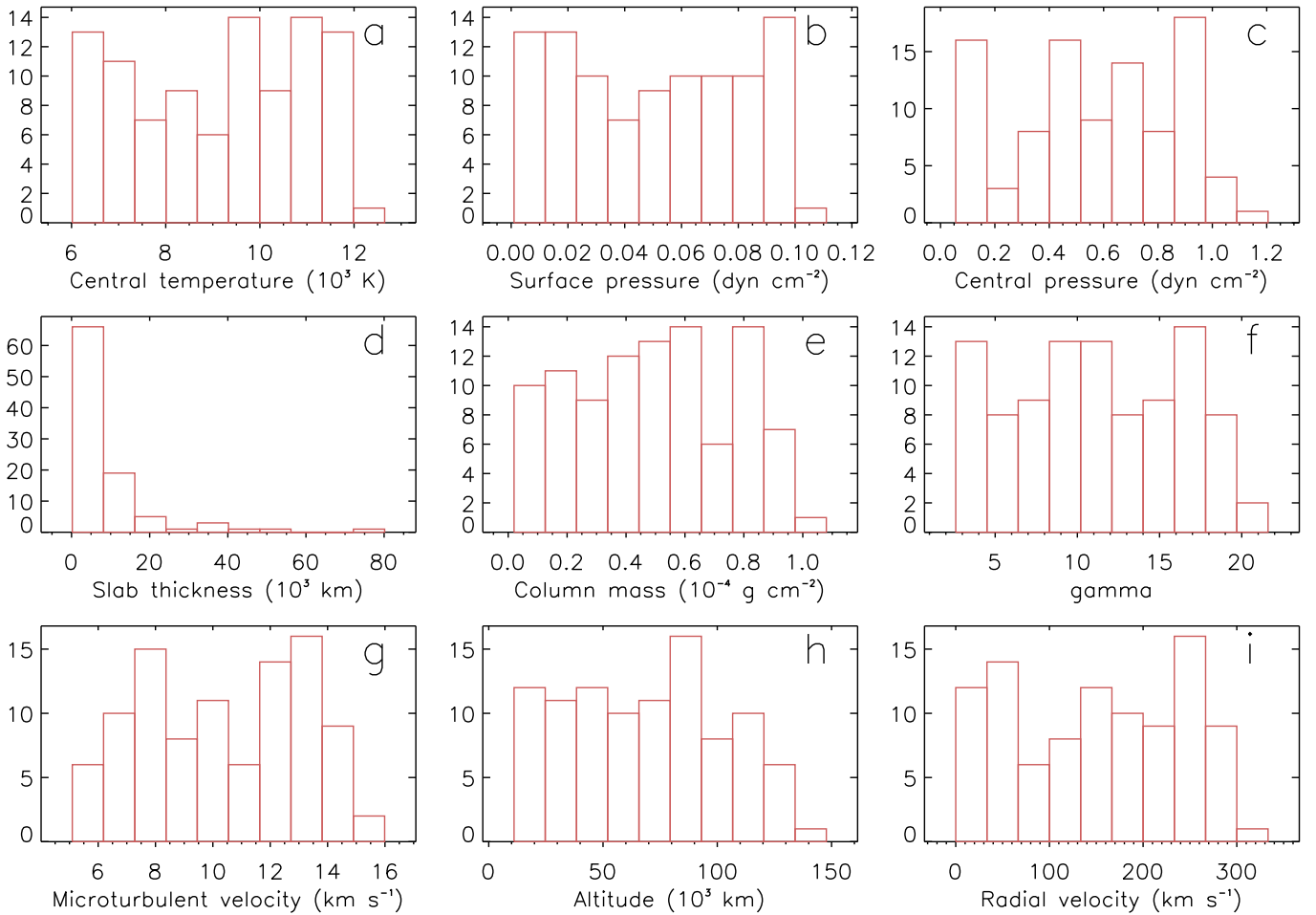}}
\caption{{Histogram densities of the prominence model parameters varying as indicated in Table~\ref{t:params}. a) Central temperature (10$^3$ K); b) Surface pressure (dyn cm$^{-2}$); c) Central pressure (dyn cm$^{-2}$); d) Slab thickness (10$^3$ km); e) Column mass (10$^{-4}$ g cm$^{-2}$); f) $\gamma$; g) Microturbulent velocity (km s$^{-1}$); h) Altitude (10$^3$ km); i) Radial velocity (km s$^{-1}$).}}
\label{f:histo}
\end{figure*}

For this study we compute a new grid of models from randomly chosen input parameters within realistic ranges of values. These are indicated in Table~\ref{t:params}.
\begin{table}
\caption{Input parameters for non-LTE models.}
\label{t:params}
\centering
\begin{tabular}{cc}
\hline\hline
Parameter & Range\\
\hline
$V_\mathrm{rad}$ (km s$^{-1}$) & 0 -- 300\\
$T_\mathrm{cen}$ (K)& 6000 -- 12000\\
$p_\mathrm{cen}$ (dyn cm$^{-2}$) & 0.001 -- 1.1\\
$p_0$ (dyn cm$^{-2})$& 0.001 -- 0.1\\
$\gamma$ & 2 -- 20\\
$M$ (g cm$^{-2}$)& $10^{-6}$ -- $10^{-4}$\\
$\xi$ (km s$^{-1}$) & 5 -- 15\\
$H$ (km) & 10000 -- 140000\\
\hline
\end{tabular}
\tablefoot{The parameters given above are, from top to bottom: radial velocity of the moving prominence, central temperature, central pressure, boundary pressure, $\gamma$ (indicating the steepness of the temperature profile), column mass, microturbulent velocity, and altitude. All parameter values are randomly chosen within the given range. The temperature at the boundary with the corona is fixed to $10^5$~K.}
\end{table}
We calculated 100 models with these input parameters.
{The realistic range of values taken by our input parameters allows us to investigate the effect of the variation of the plasma parameters on the evolution of the emergent intensity of the \ion{He}{ii} 304 line in different types of eruptive prominence models. The histogram densities for all input parameters are shown in Fig.~\ref{f:histo}.}
{There is a reasonable number of models covering all possible values of the plasma parameters. In this study, the slab width is determined from the solution of the hydrostatic equilibrium, with the column mass, the temperature and the pressure as input.}
For each model, we solve the radiative transfer and statistical equilibrium equations for Hydrogen and Helium to obtain the emergent intensities in the \ion{He}{ii} 304~\AA\ line.
For more details on theoretical investigations of the Doppler dimming and brightening effects on active and eruptive prominences, see the review by \cite{2010SSRv..151..243L}.

\subsection{Results}

{Within our randomly generated grid of models, we look for the model that has the lowest possible velocity.}
This model has the following characteristics: $T_\mathrm{cen}\approx8800$~K, pressures of 0.7 and 0.086~dyn cm$^{-2}$ at the centre of the slab and at the coronal boundary respectively, and a column mass of 4.8$\times10^{-5}$~g cm$^{-2}$ {(slab thickness of 2500~km)}. The mean temperature of the model across the slab is 11300~K.
{Its radial velocity is null. With these parameters, we can consider that this model is representative of a random quiescent prominence, and we therefore call it our reference model.}
{Starting from this reference model, we compute ten additional models where all the plasma parameters are kept constant and identical to the parameters of the reference model described above, with the radial velocity being the only variable changing. Hence we are able to clearly identify how the Doppler dimming effect alone affects the emergent intensity of the \ion{He}{ii} 304 line for our reference model.}

\begin{figure}
\resizebox{\hsize}{!}{\includegraphics{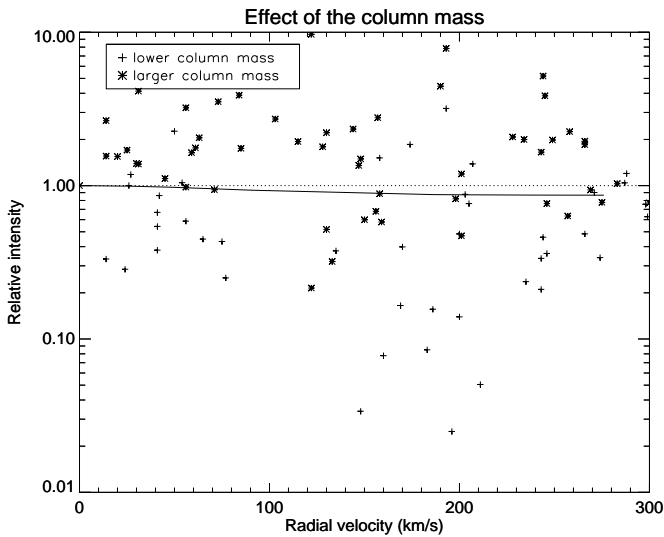}}
\caption{Effect of the column mass of the model on the relative intensities of the \ion{He}{ii} 304~\AA\ line (normalised to the intensity of {the reference} model where the prominence is at rest) as a function of the radial velocity of the prominence. {The solid line shows the variation of intensity with radial velocity when all the model parameters are kept constant.}}
\label{f:cm}
\end{figure}

Our results {(shown in Fig.~\ref{f:cm})} indicate that when all the plasma parameters are allowed to vary, {models with a non-zero velocity yield in equal proportions computed intensities in the \ion{He}{ii} line either greater or lower than the intensity of the reference (static) model. }
As our aim here is to compare our theoretical results with the observations presented in the previous section, and also to identify under what conditions an erupting prominence might become dimmer or brighter as its radial velocity increases, we look at relative intensities in the 304 line, i.e. intensities normalised to the intensity of our reference model {(which has a null radial velocity)}.

As was already pointed out by \cite{2008AnGeo..26.2961L} in cases where a PCTR is included in the modelling, the Doppler dimming effect will be stronger for models with a low column mass. This is explained by the fact that, as the column mass increases, the amount of material at high temperatures becomes more important. The sensitivity of this high temperature plasma to Doppler dimming is less strong due to the enhanced role of thermal processes in the formation of the line.
We find that among the dimmer models, 67\% have a lower column mass than the reference model (Fig.~\ref{f:cm}).
On the other hand, among the brighter models, 79\% have a larger column mass than the reference model. Hence, with these PCTR models, our results indicate that the main physical parameter giving a hint of whether the intensity at 304~\AA\ will decrease or increase with radial velocity is the column mass. 
If the column mass increases during the eruption process, for instance as a result of mass loading or rearrangement of the prominence plasma along the line of sight \citep[due, e.g., to the rotation of the filament structure -- see][]{1987SoPh..108..251K,2011A&A...531A.147B}, there is a high probability that the intensity will increase {despite an increase of the} radial velocity.

Another factor that will determine the relative importance of scattering of the incident radiation \emph{versus} collisional processes in the formation of the \ion{He}{ii} 304~\AA\ line is the temperature of the plasma.
The majority (62\%) of models with a larger column mass than our reference model will have a lower emergent intensity at 304~\AA\ in a moving prominence if the central temperature of the slab is lower than that of the reference model. Similarly, the majority of these higher column mass models are brighter at a non-zero velocity if their central temperature is larger.

These results indicate that the behaviour of the \ion{He}{ii} 304 line intensity with velocity changes according to the plasma parameters inside the prominence, in particular the column mass of the plasma and its temperature.

\section{Discussion}\label{s:discuss}

\subsection{Contributing ions in the 304~\AA\ channel}\label{s:contrib}

One difficult aspect of interpreting observations in the 304 channel is that the \ion{He}{ii} line at 303.8~\AA\ is not the only contributor. The \ion{Si}{xi} line at 303.3~\AA\ {(formed at $\sim$1.2~MK)} also contributes in the filter spectral band.

In the context of AIA data, it is believed that the \ion{Si}{xi} line can contribute to up to 20\% of the observed intensity in an active region on the disk \citep{2011SoPh..tmp..115L}. Based on results obtained by the SOHO/CDS spectrometer, \cite{2000SoPh..195...45T} also point out that the relative contribution of \ion{Si}{xi} 303.3~\AA\ can be larger than 90\% above the limb {in a typical region of the quiet corona}. However, {typical} prominence conditions are much more akin to {on-disk} quiet-sun conditions, for which the contribution of the \ion{Si}{xi} line is {reduced to} around 4\% {\citep{2000SoPh..195...45T}}.
This aspect has also been recently discussed by \cite{2011JASTP..73.1117B} in the context of STEREO/EUVI observations.
{There will also be a contribution from \ion{Si}{xi} from the corona in front of the prominence. However, it can be neglected since the observed prominences exhibit a rather sharp edge (\textit{i.e.,} the corona surrounding the prominence is much dimmer).}

While the relative contributions of these two lines will vary according to the conditions inside the plasma, and in the absence of detailed line profiles, 
{we can safely assume based on the above discussion that in prominences,}
the main contributor in
{the 304} channel is the \ion{He}{ii} line.

\subsection{Projection effects}

Another observational limitation is the lack of three-dimensional information on the eruptive structure.
We are led to make the assumption that all motions take place in the plane of the sky, and that motions take place along the radial direction.
{A possible approach}
to overcome this limitation is to use spectroscopic data to measure Doppler shifts along high-cadence imaging {to measure transverse motions}.

\cite{2011JASTP..73.1117B} mentions some of the most successful attempts at characterising the 3D motion of eruptive prominences. As found by, e.g., \cite{2010SoPh..267...95Z}, the trajectories of eruptive prominences are generally not purely in the radial direction.
However, we do not have enough information with the available data set used in this paper to refine our radial velocity estimates.
What really matters in determining the Doppler dimming / brightening effect is the radial velocity of the plasma, and we will assume that the velocities measured from the observed motions in the plane of the sky are reasonable estimates of the radial velocity of the erupting structure.
Our derived values are in agreement with values given, e.g., by \cite{2011A&A...533L...1R,2007ApJ...663.1354K}.

{In a future study, we plan to make a quantitative analysis of one particular event, studying its evolution and determining the velocity of the plasma in 3D using additional data from the STEREO spacecrafts \citep{2009Icar..200..351T,2011ApJ...730..104J,2011A&A...531A.147B}, and comparing calibrated intensities with our radiative transfer calculations.}

\subsection{Local prominence plasma conditions}

The fact that our computations can explain an increase in intensity with increasing radial velocity does not contradict the fact that the \ion{He}{ii} 304~\AA\ line is strongly sensitive to the Doppler \emph{dimming} effect.
{The effect of Doppler dimming alone (\textit{i.e.,} the variation of intensity with velocity as all plasma parameters are kept constant) is shown by the solid curve in Fig.~\ref{f:cm}. The sensitivity of the emergent intensity to Doppler dimming is explained by the predominance of resonant scattering of incoming photons from the Sun in the formation of the \ion{He}{ii} 304~\AA\ line.}
{This Doppler dimming effect (diminution of emergent intensity) is due to the fact}
that the incident radiation
becomes out of resonance with the absorption profile of the
{304~\AA}
transition as the velocity increases.
{However, it is possible that}
changes in the physical conditions of the prominence plasma
{will more than compensate the Doppler dimming effect, implying}
that a higher {emergent} intensity will be observed.

We
{looked at}
the AIA data for these four events at other wavelengths to seek signs of temperature variations of the prominence region which was tracked.
In particular, all four eruptive prominences investigated here show up in some degree in the 171~\AA\ channel (which has quite a broad temperature response), especially for the 2011-06-10 event in which the structure is very well preserved (see Sect.~\ref{s:100611}).
For three of these events, as time elapses it is only the prominence edges that remain apparent in the 171~\AA\ waveband. Only the
2011-06-10 eruptive prominence remains fairly well seen for most of the event at this wavelength.
This {could be} interpreted as evidence for localised heating of the plasma during the eruption.

\section{Conclusions}\label{s:concl}

The first detailed computations of the helium spectrum in a quiescent prominence were obtained by \cite{2001A&A...380..323L,2004ApJ...617..614L}. They showed that the ionised helium resonance lines were mostly formed by scattering of the incident radiation under typical quiescent prominence conditions. This was confirmed in a recent observational study by \cite{2011A&A...531A..69L}. As a consequence it is natural to expect that active and eruptive prominences with pronounced radial outflows
are very sensitive to the Doppler dimming effect in these lines, as demonstrated by \cite{2007A&A...463.1171L,2008AnGeo..26.2961L}.

This paper
{investigates}
the intensity variations {as a function of velocity} in eruptive prominences observed with the SDO/AIA imager in the 304 channel.
{We study four well-observed eruptive prominences and find three cases where the intensities of small features inside the eruptive structures decrease with increasing velocity, while in the fourth case, the intensities of the small features increase with velocity.}
{We then compare}
our observations with a grid of
{randomly generated} non-LTE models
{allowing us to study the effects of the plasma parameters and of the radial velocity of the prominence plasma on the emergent intensity of the \ion{He}{ii} 304~\AA\ line.}
{Our theoretical results confirm that the intensity of the \ion{He}{ii} line can increase as well as decrease when the velocity of the prominence plasma increases and the model plasma parameters are allowed to vary.}

{The answer to the question of whether the emergent intensity in the \ion{He}{ii} 304 line decreases or increases with increasing radial velocity of the eruptive prominence depends on the variations of the local plasma conditions as well as on the velocity of the prominence. Even if previous theoretical studies showed that this line is strongly sensitive to the Doppler dimming effect, variations in the plasma parameters also play an important role in determining the behaviour of the emergent intensity.}
{Specifically}, the column mass and the temperature of the plasma {are found to be the two most important parameters in determining the evolution of the intensity of the 304~\AA\ line with velocity of the prominence plasma}.

{This} non-LTE modelling {of eruptive prominences opens the door for improved diagnostics, allowing us to}
{shed light on the variations of}
the plasma parameters of eruptive prominences from SDO/AIA observations at 304~\AA.
This diagnostic technique, applied in the present study to SDO/AIA observations, can be applied to any other past \citep[EIT on SOHO,][]{1995SoPh..162..291D}, current \citep[EUVI, part of SECCHI on STEREO,][]{2008SSRv..136...67H}, or future EUV observations of prominence eruptions observed at a sufficiently high cadence in the \ion{He}{ii} 304~\AA\ channel.

In a future study we plan to make a more detailed diagnostic of the eruptive prominence plasma by comparing absolute intensities in the 304~\AA\ line with a suitable grid of non-LTE models.

\begin{acknowledgements}
{We thank the referee for helpful suggestions to improve the clarity of the paper.}
Part of this work was supported by a Nuffield Foundation Undergraduate Research Bursary to KM (URB/39921).
Financial support by the International Space Science Institute (ISSI) is gratefully acknowledged, and NL thanks colleagues from the ISSI International Team 174 on "Solar Prominence Formation and Equilibrium: New Data, New Models" for stimulating discussions.
{The data used are provided courtesy of NASA/SDO and the AIA science team. We thank Greg Slater for his help with the processing of the Level 1 AIA data.}
This research has made use of NASA's Astrophysics Data System.
\end{acknowledgements}

\bibliographystyle{aa} 

\end{document}